\documentclass[12pt]{article}

\usepackage{amsmath, amssymb,graphicx, pstricks,pst-coil,pst-plot,psfrag,bbm }

  \newcommand{\N}{\mathcal{N}}

\newcommand{\beq}{\begin{equation}}
\newcommand{\eeq}{\end{equation}}
\newcommand{\beqa}{\begin{eqnarray}}
\newcommand{\eeqa}{\end{eqnarray}}
\newcommand{\beqar}{\begin{eqnarray*}}
\newcommand{\eeqar}{\end{eqnarray*}}

\newcommand{\p}{\phi}

\def\p1{\phantom{1}}
\def\IR{{\hbox{{\rm I}\kern-.2em\hbox{\rm R}}}}

\newcommand{\CC}{\mathbb{C}}
\newcommand{\RR}{\mathbb{R}}
\newcommand{\ZZ}{\mathbb{Z}}
\newcommand{\PP}{\mathbb{P}}

\newcommand{\cp}{{\mathbb{CP}^3}}
\newcommand{\rp}{{\mathbb{RP}^3}}

\newcommand{\supai}[1]{a_{#1} }

\newcommand{\supbi}[1]{b_{#1} }

\newcommand{\supqi}[1]{q_{#1} }

\newcommand{\suptqi}[1]{\tilde{q}_{#1} }

\newcommand{\dq}[2]{q_{#2}^{#1}}
\newcommand{\bdq}[2]{\bar{q}^{#2}_{#1}}

\newcommand{\dua}[2]{a_{#2}^{#1}}
\newcommand{\bdua}[2]{\bar{a}^{#2}_{#1}}

\begin{document}

%%%%%%%%%%%%%%%%%%%%%%%%%%%%   TITLE    %%%%%%%%%%%%%%%%%%%%%%%%%%%%%%%%%

\thispagestyle{empty}
\renewcommand{\thefootnote}{\fnsymbol{footnote}}

\mbox{}
\bigskip\bigskip\bigskip

\begin{center} \noindent \Large \bf A note on the holography of \\ 
Chern-Simons  matter theories with flavour
\end{center}

\bigskip\bigskip\bigskip \centerline{ Stefan Hohenegger and Ingo
  Kirsch$^a$\footnote[1]{\noindent \tt email:
    stefanh@phys.ethz.ch,\\\mbox{}\hspace{1.91cm} kirsch@phys.ethz.ch} }

\bigskip\bigskip

\centerline{\it ${}^a$ Institut f\"ur Theoretische Physik, ETH
  Z\"urich}
\centerline{\it 
CH-8093 Z\"urich, Switzerland}
\bigskip\bigskip

\bigskip\bigskip

\renewcommand{\thefootnote}{\arabic{footnote}}

\centerline{\bf \small Abstract}
\medskip

{\small \noindent We study a three-dimensional $\N=3$ $U(N)_k \times
  U(N)_{-k}$ Chern-Simons matter theory with flavour, corresponding to
  the $\N=6$ Aharony-Bergman-Jafferis-Maldacena CSM theory coupled to
  $2N_f$ fundamental fields. The dual holographic description is given
  by the near-horizon geometry of $N$ M2-branes at a particular
  hypertoric geometry ${\cal M}_8$. We explicitly construct the space
  ${\cal M}_8$ and match its isometries to the global symmetries of
  the field theory.  We also discuss the model in the quenched
  approximation by embedding probe D6-branes in $AdS_4 \times \cp$.}

\newpage

%%%%%%%%%%%%%%%%%%%%%%%%%%%%%%%%%%%%%%%%%%%%%%%%%%%%%%%%%

\setcounter{equation}{0}
\section{Introduction}

Recently, there has been a renewed interest in three-dimensional
superconformal Chern-Simons-matter (CSM) theories. Other than their
purely topological cousins, this type of Chern-Simons theories
exhibits non-trivial dynamics due to the coupling to matter fields.
Bagger and Lambert \cite{BL} as well as Gustavsson \cite{G} (BLG)
constructed a three-dimensional $\N=8$ superconformal Chern-Simons
gauge theory with manifest $SO(8)$ R-symmetry. A unitary realization
of the involved three-algebra restricted the gauge group to $SO(4)$.
After the reformulation of the BLG theory as a $SU(2) \times SU(2)$
CSM theory \cite{Raamsdonk}, Aharony, Bergman, Jafferis and Maldacena
(ABJM) \cite{ABJM} constructed a $\N=6$ CSM theory with gauge group
$U(N) \times U(N)$ at level~$k$ as the world-volume theory of $N$
M2-branes at a $\CC^4/\ZZ_k$ orbifold.

A prerequisite for making the above theory interesting for more
realistic applications, e.g.\ in condensed matter physics, is the
introduction of light matter fields in the {\em fundamental}
representation of the gauge group. The fundamentals could serve, for
instance, as a prototype for strongly-coupled electrons.  First steps
in this direction have been taken in \cite{Giveon2008, Niarchos,
  Niarchos:2009aa}, which discussed $\N=2$ supersymmetric CSM theories
with fundamental matter and discovered an interesting strong-weak
coupling Seiberg-type duality.  However, Refs.~\cite{Giveon2008,
  Niarchos, Niarchos:2009aa} have not yet addressed a possible
holographic description of CSM theories with flavour, which over the
last years has turned out to be remarkably successful for Yang-Mills
theories (see e.g.~\cite{Erdmenger:2007cm} for a review).

In this note we fill this gap by proposing a holographic description
of the ABJM model coupled to $2N_f$ light fundamental fields. We show
that the field theory, whose action will be written using $\N=2$
superspace formalism, preserves $\N=3$ supersymmetry for particular
values of the coupling constants. We find that other than in the
(unflavoured) ABJM model, where supersymmetry is enhanced to $\N=6$
\cite{ABJM}, the supersymmetry of the present model remains $\N=3$ in
the infrared. The latter describes the low-energy region of the
open-string sector of the web-deformed type~IIB configuration studied
in \cite{ABJM} with two additional stacks of $N_f$ D5-branes. The
T-dual type~IIA setup, now involving $2N_f$ D6-branes, lifts to $N$
M2-branes at the origin of a toric hyperk\"ahler geometry ${\cal
  M}_8$. We explicitly construct ${\cal M}_8$ and compare its isometry
group to the global symmetries of the dual $\N=3$ field theory.

The corresponding near-horizon geometry includes the information of
the (uplifted) flavour D6-branes and therefore their backreaction on
the geometry. However, the complicated structure of the near-horizon
metric impedes further progress along these lines.  We therefore
continue to discuss flavours in the quenched approximation, using
holographic methods as initiated in \cite{KarchKatz, Kruczenski,
  Babington}. This requires the embedding of probe D6-branes in $AdS_4
\times \cp$, which is the near-horizon geometry of the ABJM setup in
type-IIA string theory \cite{ABJM}. The D6-branes fill the $AdS_4$
space and wrap around a special Lagrangian submanifold inside the
$\cp$.  We show that the real projective space $\RR\PP^3$ is such a
submanifold inside the $\cp$, and thus the corresponding embedding of
the D6-branes is stable and supersymmetric.

The paper is organized as follows. In section~\ref{secft} we present
the $\N=3$ Chern-Simons Yang-Mills theory with matter in the
fundamental representation of the $U(N)_k \times U(N)_{-k}$ gauge
group. In section~\ref{secsetup} we discuss the corresponding brane
setup in type IIB string theory, its lift to $M$-theory and the
corresponding near-horizon geometry.  In section \ref{secprobe} we
discuss the embedding of probe D6-branes in $AdS_4 \times \CC\PP^3$.

\medskip {\bf Note added:} After publication of the first version of
this work, two further papers \cite{Gaiotto2, Hikida} appeared in the
arXiv, which have considerable overlap with the present work. In
particular, taking into account a comment in the introduction of
\cite{Gaiotto2}, we clarified the discussion of our brane-setup in
section~3.

%%%%%%%%%%%%%%%%%%%%%%%%%%%%%%%%%%%%%%%%%%%%%%%%%%%%%%%%%%%%%
\setcounter{equation}{0}
\section{Chern-Simons Yang-Mills theory with fundamental matter}
\label{secft}

In this section we study a three-dimensional $\N=3$ superconformal
$U(N)\times U(N)$ Chern-Simons-matter theory with flavour in the
fundamental representation of the gauge group. This theory will be
obtained by coupling $N_f$ fundamental hypermultiplets to the ABJM
theory~\cite{ABJM}.

\subsection{The action}
The ABJM theory has gauge group $U(N)_k\times U(N)_{-k}$ and its action can
be written in manifest $\N=2$ language~\cite{ABJM}. Let us briefly
review its field content. There are two bifundamental $\N=4$
hypermultiplets $(A, B^\dagger)_{1,2}$ and two adjoint $\N=4$ vector
multiplets consisting of the $\N=2$ vector fields $V_{1,2}$ and the
chiral fields $\Phi_{1,2}$. Formally, there are also $k$ chiral
multiplets ($q_{1,2}$) in the fundamental and $k$ chiral multiplets
($\tilde q_{1,2}$) in the anti-fundamental representation of each
gauge group. These are assumed to be massive and, when integrated out,
produce a Chern-Simons term via the parity anomaly.  Thus at low
energies all fundamental fields are integrated out, leaving only
fields in the adjoint or bifundamental representation. In order to
also have massless fundamental fields in the far infrared, we
introduce $2N_f$ fundamental hypermultiplets $(Q^r, \tilde
Q^r{}^\dagger)_{1,2}$ with $r=1,...,N_f$. The $\N=2$ superfields and
their quantum numbers are summarized in the upper part of
table~\ref{table1}.

\begin{table}[!ht] 
\begin{center}
\begin{tabular}{cccccccc}
 & $U(N)$ & $U(N)$  & $U(k)$ & $U(k)$ & $U(N_f)$ & $U(N_f)$ & $\Delta$\\  
\hline
\parbox{0.45cm}{\vspace{0.1cm}$A_1$\vspace{0.1cm}}, \parbox{0.45cm}{\vspace{0.1cm}$A_2$\vspace{0.1cm}} & \parbox{0.35cm}{\vspace{0.1cm}$N$\vspace{0.1cm}} &\parbox{0.35cm}{\vspace{0.1cm}$\overline{N}$\vspace{0.1cm}} &\parbox{0.25cm}{\vspace{0.1cm}\bf{1}\vspace{0.1cm}}&\parbox{0.25cm}{\vspace{0.1cm}\bf{1}\vspace{0.1cm}}&\parbox{0.25cm}{\vspace{0.1cm}\bf{1}\vspace{0.1cm}}&\parbox{0.25cm}{\vspace{0.1cm}\bf{1}\vspace{0.1cm}}&\parbox{0.25cm}{\vspace{0.1cm}$\frac{1}{2}$\vspace{0.1cm}} \\
 $B_1$, $B_2$ & $\overline {N}$&$N$&\bf{1}&\bf{1}&\bf{1}&\bf{1} &$\frac{1}{2}$\\
 $\Phi_1$, $V_1$ & adjoint & \bf{1} & \bf{1}& \bf{1}&\bf{1}&\bf{1}&$1, 0$\\
 $\Phi_2$, $V_2$ & \bf{1} & adjoint & \bf{1}& \bf{1}&\bf{1}&\bf{1}&$1, 0$\\
\hline
\parbox{0.3cm}{\vspace{0.1cm} $q_1$\vspace{0.1cm}}  & \parbox{0.35cm}{\vspace{0.1cm}$N$\vspace{0.1cm}}   & \parbox{0.25cm}{\vspace{0.1cm}\bf{1}\vspace{0.1cm}} &\parbox{0.25cm}{\vspace{0.1cm}\bf{1}\vspace{0.1cm}}&\parbox{0.25cm}{\vspace{0.1cm}$\overline{k}$\vspace{0.1cm}}&\parbox{0.25cm}{\vspace{0.1cm}\bf{1}\vspace{0.1cm}}&\parbox{0.25cm}{\vspace{0.1cm}\bf{1}\vspace{0.1cm}}   &\parbox{0.25cm}{\vspace{0.1cm}$\frac{1}{2}$\vspace{0.1cm}}\\
 $\tilde q_1$  & $\overline {N}$&\bf{1}&$k$&\bf{1}&\bf{1}&\bf{1} &$\frac{1}{2}$\\
 $q_2$  & \bf{1} & $N$  &$\overline{k}$ & \bf{1}&\bf{1}&\bf{1}&$\frac{1}{2}$\\
 $\tilde q_2$  & \bf{1} &$\overline {N}$& \bf{1}& $k$&\bf{1}&\bf{1}&\parbox{0.25cm}{\vspace{0.1cm}$\frac{1}{2}$\vspace{0.1cm}}\\
\hline
 \parbox{0.45cm}{\vspace{0.1cm}$Q_1$\vspace{0.1cm}}, \parbox{0.45cm}{\vspace{0.1cm}$\tilde Q_1^\dagger$\vspace{0.1cm}}&$N$&\bf{1}&\bf{1}&\bf{1}&$N_f$&\bf{1}
 &$\frac{1}{2}$\\
 $Q_2$, $\tilde Q_2^\dagger$&\bf{1}&$N$&\bf{1}&\bf{1}&\bf{1}&$N_f$
 &$\frac{1}{2}$\\
\end{tabular}
\end{center}
\caption{$\N=2, d=3$ superfields  in the field theory.}\label{table1}
\end{table}

%\noindent
In $\N=2$ superspace, the action can be written as a sum of three
terms ${\cal S}={\cal S}_{\text{mat}}+ {\cal S}_{\text{CS}}+{\cal
  S}_{\text{pot}}$, a matter part, a Chern-Simons part and a
superpotential given by
\begin{align}
{\cal S}_{\text{mat}}&= \int d^3 x d^4 \theta {\,\rm Tr} \left(-\bar A_i e^{-V_1} A_i e^{V_2} - \bar B_i e^{-V_2} B_i e^{V_1}\right) - \bar Q^r_i e^{-V_i} Q^r_i  -  {\tilde Q}^r_i e^{V_i} \bar{\tilde Q}^r_i \,,\label{ActPart1}\\  
{\cal S}_{\text{CS}}&= -i \frac{k}{4\pi} \int d^3 x d^4 \theta \int_0^1 dt {\,\rm Tr}\left(V_1 \bar D^\alpha(e^{t V_1} D_\alpha e^{-tV_1})- V_2 \bar D^\alpha(e^{t V_2} D_\alpha e^{-tV_2}) \right)\,,\label{ActPart2}\\
{\cal S}_{\text{pot}} &=\int d^3 x d^2 \theta\, \left(W_{\text{ABJM}} +  W_{\text{flavour}}\right) + c.c.  \,,\label{ActPart3} 
\end{align}
where 
\begin{align}\label{WABJM}
W_{\text{ABJM}} =-\frac{k}{8\pi}{\,\rm Tr\,} (\Phi_1^2-\Phi_2^2) 
+ {\,\rm Tr\,} (B_i \Phi_1 A_i)
+ {\,\rm Tr\,} (A_i \Phi_2 B_i)
\end{align}
and 
\begin{align} \label{Wflavor}
W_{\text{flavour}} = \alpha_1\tilde Q^r_1 \Phi_1 Q^r_1 +\alpha_2 \tilde Q^r_2 \Phi_2 Q^r_2  \,.
\end{align}
The first term in the ABJM superpotential $W_{\text{ABJM}}$
\cite{ABJM} involving $\Phi^2_1$ and $\Phi^2_2$ is the $\N = 3$
supersymmetry completion of the Chern-Simons action ${\cal
  S}_{\text{CS}}$, while the remaining two terms include the coupling
to the bifundamentals $A_{1,2}$ and $B_{1,2}$.  The superpotential
$W_{\text{flavour}}$ describes the coupling of the new flavour fields
$\tilde Q^r_{1,2}, Q^r_{1,2}$ to the adjoints $\Phi_{1,2}$. The action
preserves $\N=2$ supersymmetry for arbitrary values of the coupling constants $\alpha_{1,2}$.  

\begin{figure}[ht]
\begin{center}
\input{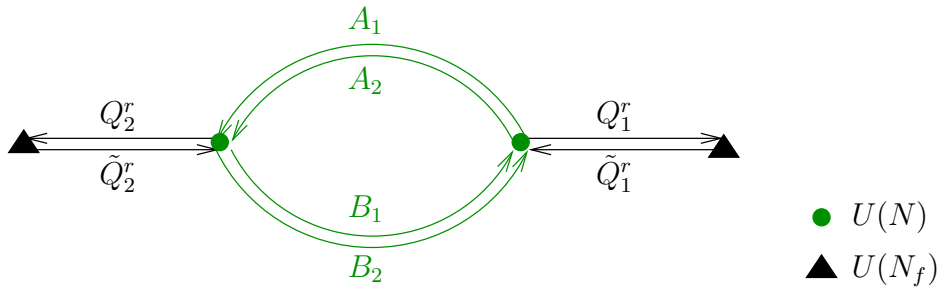}
\caption{Quiver diagram of the Chern-Simons-Yang-Mills theory with flavours.}
\label{quiverfig}
\end{center}
\end{figure}

There are no kinetic terms for the fields of the $\N=4$ vector
multiplet, which contains the $\N=2$ superfields $V_{1,2}$ and
$\Phi_{1,2}$. These fields are massive and will be integrated out at
low energies. Upon integrating out the adjoint fields $\Phi_{1,2}$, we
get the superpotential
\begin{align}
W&=W_{\text{ABJM}} + W_{\text{flavour}} \nonumber \\
&= \frac{4\pi}{k} {\,\rm Tr\,} (A_1B_1A_2B_2-A_2B_1A_1B_2)
+\frac{4\pi\alpha_1}{k}\,\tilde Q_1(A_1B_1+A_2B_2)Q_1\nonumber\\
&\quad-\frac{4\pi\alpha_2}{k}\,\tilde Q_2 (B_1 A_1 +B_2A_2) Q_2+
\frac{2\pi\alpha_1^2}{k}\, Q_1 \tilde Q_1 Q_1 \tilde
Q_1-\frac{2\pi\alpha_2^2}{k}\,\tilde Q_2Q_2 \tilde Q_2 Q_2 \,.
\label{superpot}
\end{align}
The first term is exactly the same as in the Klebanov-Witten theory
associated with the conifold \cite{conifold}. The remaining terms
proportional to $\alpha_1$ and $\alpha_2$ describe the coupling of the
fundamentals to the ABJM model. Similar terms appear when fundamental
matter is coupled to the Klebanov-Witten theory, see for instance
\cite{Kuperstein, Ouyang}.  The field content and the superpotential
of the low-energy theory can best be represented by the quiver diagram
shown in figure \ref{quiverfig}.

%%%%%%%%%%%%%%%%%%%%%%%%%%%%%%%%%%%%%%%%%%%%%%%

\subsection{$\N=3$ supersymmetric theory and conformal invariance}
So far we have considered generic coupling constants $\alpha_1$ and
$\alpha_2$. However, it turns out that upon choosing the particular
values $\alpha_1=-\alpha_2=1$, the amount of supersymmetry preserved
by the action (\ref{ActPart1})--(\ref{ActPart3}) is enhanced to
$\N=3$. This is accompanied by an enhancement of the $U(1)_R$
R-symmetry to $SU(2)_R$, which is explicitly shown in
appendix~\ref{Sect:InvCompAct}, where we write the bosonic part of
the action in a manifest $SU(2)_R$ invariant way.

As we have also shown in appendix \ref{Sect:InvCompAct}, apart from
the $SU(2)_R$ symmetry, the action is also invariant under an
additional $SU(2)_D$ symmetry.\footnote{This symmetry simultaneously
  exchanges $A_1$ with $A_2$ and $B_1$ with $B_2$ and can be thought
  of as the diagonal $SU(2)$ of the global $SU(2)_A \times SU(2)_B$
  group of the ABJM model \cite{ABJM}.} It is important to notice that
the latter is a global symmetry which, in particular, commutes with
$SU(2)_R$. Therefore, there is no enhancement of the R-symmetry group
(or supersymmetry), in contrast to the ABJM model \cite{ABJM} and
related theories, e.g.~\cite{Benna, Jafferis}.

In addition to the $SU(2)_R \times SU(2)_D$ symmetry of
(\ref{superpot}) there is finally also the ``baryonic'' $U(1)$
symmetry
\begin{align}
&U(1)_b: && A_i \rightarrow e^{i \alpha} A_i \,,&&
 B_i \rightarrow e^{-i \alpha} B_i \,,&& Q^r_i, \tilde Q^r_i \,\,\,
 \textmd{inert} \,, \label{baryonU1}
\end{align}
which has already been discussed in detail in \cite{ABJM}. This
symmetry has to be distinguished from the baryonic $(U(1) \times
U(1))_{B}$ subgroup of the $U(N_f) \times U(N_f)$ flavour group.

\medskip All couplings in (\ref{superpot}) are marginal, and the
theory is classically conformal \mbox{invariant}. The standard
non-renormalization theorem for 2+1-dimensional Yang-Mills theories
coupled to matter fields does not apply to CSM theories
\cite{Gaiotto}. Nevertheless, there are good reasons to believe that,
similarly to the ABJM theory \cite{ABJM} and the general class of CSM
theories studied in \cite{Gaiotto}, the present $\N=3$ CSM theory
($\alpha_1=-\alpha_2=1$) is also conformal invariant at the quantum
level.  Note first that the Chern-Simons level~$k$ is not renormalized
beyond a possible one-loop shift \cite{Kapustin:1994mt}. Moreover, as
found in~\cite{Gaiotto}, possible corrections to the classical
K\"ahler potential are either irrelevant or absorbed by a wave
function renormalization. However, since for $\N=3$ supersymmetry,
$U(1)_R$ is part of the (non-anomalous) $SU(2)_R$ R-symmetry, the
conformal dimensions of all fields are protected from quantum
corrections.  Therefore, there is no $U(1)_R$ charge renormalization
nor wave function renormalization, excluding relevant or marginal
corrections to the K\"ahler potential.  Non-renormalization of the
coupling constants in the superpotential has explicitly been shown to
two-loop order in \cite{Avdeev} for CSM theories with matter fields in
the fundamental representation. We expect that the coupling to the
ABJM term does not destroy the non-renormalization. This strongly
suggests conformal invariance of the action at the quantum level.

%%%%%%%%%%%%%%%%%%%%%%%%%%%%%%%%%%%%%%%%%%%%%%%%%%%%%%%%%%%%%%%%%
\setcounter{equation}{0}
\section{Brane construction} \label{secsetup} 
\begin{figure}[!ht]
\begin{center}
\includegraphics[scale=0.8]{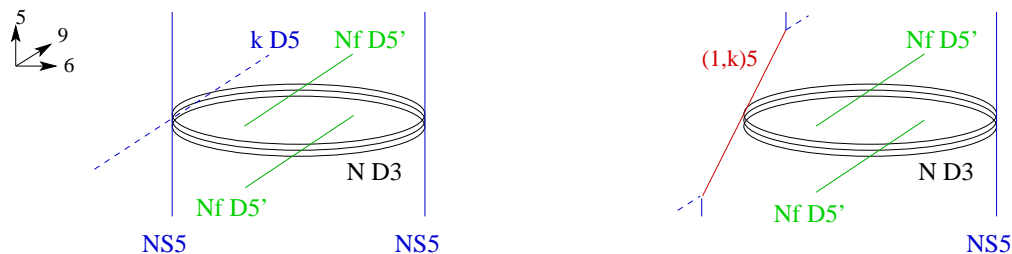}
\caption{Type IIB brane setup of \cite{ABJM} plus two stacks of $N_f$
  ``flavour'' D5-branes before (lhs.) and after the web deformation
  (rhs.). }
\label{setup}
\end{center}
\end{figure}

In this section we will
make a proposal for the gravitational theory dual to the Chern-Simons
Yang-Mills theory with fundamental matter discussed in the previous
section. The gravitational theory corresponds to the near-horizon
geometry of the following brane-construction. We start from the type
IIB setup of \cite{ABJM} which consists of two NS5-branes along
012345, which are separated in the compact direction 6, and $N$
D3-branes along 0126.  In addition, there are $k$ D5-branes along
012349 which intersect the D3-branes along 012 and one of the two
NS5-branes along 012345, as shown in figure~\ref{setup}.

This setup preserves $\N=2$ supersymmetry and gives rise to the
following \mbox{$\N=2$} \mbox{superfields} \cite{ABJM}: The NS5-branes
divide the D3-brane worldvolume into two intervals along~6.  The 3-3
open strings therefore give rise to two $U(N)$ $\N=4$ vector
multiplets $(V, \Phi)_{1,2}$ consisting of $\N=2$ vector and chiral
multiplets.  They also give rise to two complex bifundamental $\N=4$
hypermultiplets $(A, B^\dagger)_{1,2}$. Furthermore, we note that the
(left) NS5-brane splits the $k$ D5-branes into two stacks of $k$
half-D5-branes along the direction 9.  This phenomenon is dubbed {\em
  flavour doubling}, see {\em e.g.}\ \cite{Uranga}: Each stack of half
D5-branes gives rise to a $U(k)$ global symmetry and provides $2 k$
fundamental flavours, {\em i.e.}~$k$ flavours for each gauge factor.
At low energies the 3-5 and 5-3 open string modes therefore generate
$k$ fundamental chiral fields $q_{1,2}$, $\tilde q_{1,2}$. These
fields transform under the $U(k) \times U(k)$ global symmetry as
indicated in table~\ref{table1}. The remaining modes coming from
strings with both ends on 5-branes are assumed to be decoupled at low
energies.

We may now add another class of fundamental fields by introducing
$2N_f$ ``flavour'' D$5'$-branes along 012789.\footnote{The
  D$5'$-branes are actually grouped along $x^6$ in two stacks of $N_f$
  D$5'$-branes, one in each D3-brane sector.} These branes intersect
with the $k$ D5-branes on a three-brane along 0129 and overlap with
the D3-branes along 012. This does not break any further
supersymmetries, {\em i.e.}~the total configuration still preserves
$\N=2$. At low energies the 3-$5'$ and $5'$-3 strings give rise to $2
N_f$ additional fundamental hypermultiplets: $(Q^r, \tilde
Q^r{}^\dagger)_{1,2}$ with $r=1,...,N_f$. The corresponding $U(N_f)
\times U(N_f)$ flavour symmetry is non-chiral.

We now perform a {\em web deformation}, in which the $k$ D5-branes and
the NS5-brane merge into an intermediate $(1,\pm k)5$-brane along
$012[3,7]_{\theta_1} [4,8]_{\theta_2} [5,9]_{\theta_3}$, as explained
in detail in \cite{ABJM}. This notation means that the $(1,\pm
k)5$-brane is aligned along 012 and stretched along directions mixing
345 and 789. Only if the $\theta_{i}$ ($i=1,2,3$) are all the same and
satisfy $\tan \theta_i =k$ supersymmetry is enhanced from $\N=2$ to
$\N=3$ \cite{Kitao:1998mf, Gauntlett}.

\begin{figure}[!ht]
\begin{center}
\includegraphics[scale=0.9]{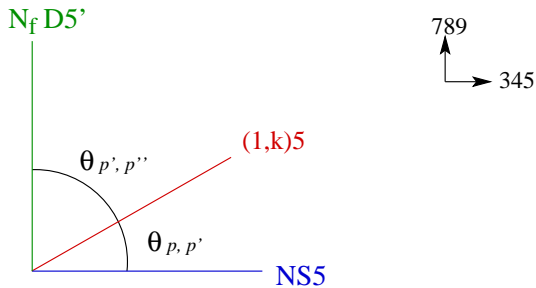}
\caption{Three types of 5-branes in $\mathbb{E}^6$.}
\label{fivebrane}
\end{center}
\end{figure}

The result of this deformation is a triple 5-brane intersection of a
$(1, 0)5$-brane (NS5-brane), a $(1, k)5$-brane, and two $(0,
N_f)5$-branes ($N_f$ D5-branes). These branes overlap over 012, and the
remaining directions of the 5-branes form three-planes in the
$\mathbb{E}^6$ parameterized by $(x^3, x^4,x^5, x^7, x^8,x^9)$. The
angle $\theta_{p, q}$ of two of these three-planes is given
by~\cite{Gauntlett}
\begin{align}
\cos \theta_{p, q} = \frac{p \cdot q}{\sqrt{p^2 {q}^2}} \,,
\end{align}
where $p\cdot q = p_i q_i$, and $p, q$ are two of the three $SL(2,
\ZZ)$ charge vectors $p=(1, 0)$, $p'=(1, k)$, $p''=(0, N_f)$. We
obtain the angles
\begin{align}
&\tan \theta_{p, p'} = k \,, &&
\tan \theta_{p', p''} = \frac{1}{k} \,,&&
\theta_{p, p''} = \frac{\pi}{2} \,,
\end{align} 
which satisfy $\theta_{p, p'} +\theta_{p', p''} = \theta_{p, p''}
=\frac{\pi}{2}$. The 5-branes and their intersection angles are shown
in figure~\ref{fivebrane}. The rotations in the three three-planes by
the same element of $SO(3)_R$ correspond to the R-symmetry
transformations of the $\N=3$ ultraviolet field theory
(\ref{ActPart1})--(\ref{Wflavor}) (with $\alpha_1=-\alpha_2=1$).

\medskip We finally note that our setup differs from those considered
in \cite{Giveon2008, Niarchos}, which study $\N=2$ supersymmetric
Chern-Simons theories with fundamental fields.  There the $(1,
k)5$-brane is rotated in the 3-7 plane, but not in the 4-8 and 5-9
plane, $\theta_1 = \theta$, $\theta_2=\theta_3=0$.  Because of this
supersymmetry is reduced to $\N=2$ there.

\subsection{T-dual setups and lift to M-theory} \label{secIIA}

As in \cite{ABJM} we begin by T-dualizing along the direction~6. The
resulting {\em web-deformed} type IIA setup then consists of the
following branes: The $N$ D3-branes map to $N$ D2-branes along 012,
and the NS5-brane turns into a single Kaluza-Klein monopole with
world-volume along 012345. The $(1, k)5$ brane is T-dual to an object
along $0126[3,7]_\theta[4,8]_\theta[5,9]_\theta$ (with $\tan
\theta=k$), which consists of $k$ D6-branes and $1$ KK monopole
associated with the 6 direction \cite{ABJM}. In addition we now have
$2N_f$ D6$'$-branes along $0126789$ descending from the flavour
D5$'$-branes in the type IIB setup.

\medskip 

The type IIA setup may now be lifted to M-theory, where the D2-branes
naturally become M2-branes along $012$. The object along
$0126[3,7]_\theta[4,8]_\theta[5,9]_\theta$ and the D6$'$-branes
become KK monopoles with circular direction in 6 and 10 and a linear
combination of both \cite{ABJM}. The resulting M-theory configuration
will be a stack of $N$ M2-branes located at the origin of a {\em toric
hyperk\"ahler manifold} \cite{Gauntlett}. This is an eight-dimensional
space ${\cal M}_8$ with $sp(2)$ holonomy and preserves $3/16$ of the
supersymmetries of the eleven-dimensional supergravity, which is
precisely the amount of supersymmetry expected for the dual of
theories in 2+1 dimensions with $\N = 3$ supersymmetry.  Adding
a~stack of $N$ M2-branes at the origin of ${\cal M}_8$ does not break any
additional supersymmetry~\cite{Gauntlett}.

The metric of ${\cal M}_8$ is given by
\begin{align}
  ds^2_{{\cal M}_8}&=U_{ij} d \vec x^i \cdot d\vec x^j + U^{ij}
  (d\varphi_i +A_i) (d\varphi_j +A_j) \,, \label{metricX8}
\end{align}
with the following quantities
\begin{align}
  &A_i = d \vec x^j \cdot \vec \omega_{ji} = d x^j_a \omega^a_{ji}
  \,,&& \partial_{x^j_a} \omega^b_{ki} - \partial_{x^k_b}
  \omega^a_{ji} = \epsilon^{abc} \partial_{x^j_c} U_{ki} \,,
  \label{relX8}
\end{align}
with $i,j=1,2$. The three-vectors $\vec x_1$ and $\vec x_2$ describe
positions in two three-planes parameterized by $(x^7,x^8,x^9)$ and $(x^3,
x^4,x^5)$, respectively. The two circular directions of the toric
geometry are in the directions 6 and 10.

The two-dimensional matrix $U_{ij}$ contains the information of the
uplifted five-branes of the IIB setup \cite{Gauntlett}. Here it is
given by
\begin{align} \label{U}
U= \mathbf{1}+\begin{pmatrix} h_1 &0\\0 & 0 \end{pmatrix}
             +\begin{pmatrix} h_2 & kh_2\\ kh_2 & k^2h_2  \end{pmatrix}
             +\begin{pmatrix} 0 &0\\0&N^2_f h_3 \end{pmatrix}\,,
\end{align}
with
\begin{align}
&h_1= \frac{1}{2|\vec x_1|}\,,
&&h_2= \frac{1}{2|\vec x_1 + k \vec x_2|}\,,
&&h_3= \frac{1}{|N_f \vec x_2|}\,.
\end{align}
The first three terms in (\ref{U}) are as in the ABJM case without
flavours \cite{ABJM}, while the last term contains the information of
the uplifted flavour branes. In the type IIB setup the functions
$h_{1,2,3}$ stem from the $(1, 0)5$-brane (NS5-brane), the $(1,
k)5$-brane, and the (two stacks of) $(0, N_f)5$-branes (D5-branes),
respectively.

An appropriate ansatz for $N$ M2-branes at the origin of ${\cal M}_8$ is
\begin{align}
ds^2 &= H^{-2/3} (-dX_0^2 + dX_1^2 +dX_2^2 )+ H^{1/3} ds_{{\cal M}_8}^2 \,,\\
   F &= dX_0 \wedge dX_1 \wedge dX_2 \wedge dH^{-1} \,,
\end{align}
where the scalar function $H$ only depends on the coordinates of
${\cal M}_8$. The supergravity equations of motion then require that
$H$ satisfies the Laplace equation on ${\cal M}_8$,
\begin{align}
\partial_\mu (\sqrt{g}g^{\mu\nu} \partial_\nu H) = 0 \,, \label{Laplace}
\end{align}
with $g_{\mu\nu}$ given by (\ref{metricX8}).

%%%%%%%%%%%%%%%%%%%%%%%%%%%%%%%%%%%%%%%%%%%%%%%%%%%%%%%%%%%
\subsection{Near-horizon geometry} \label{nhg} 
Here we do not attempt to explicitly solve (\ref{Laplace}) but instead
explore the hypertoric geometry of the manifold ${\cal M}_8$ in more
detail.  Given the form (\ref{U}) of the matrix $U_{ij}$, we see that
the metric (\ref{metricX8}) develops a physical singularity at the
point $\vec{x}_1=\vec{x}_2=0$. In this near-core region the constant
piece of the matrix (\ref{U}) is subdominant and can henceforth be
dropped.  In the following we will carry on to study this region more
closely.  We begin by presenting the solution to the equations
(\ref{relX8}) by writing an expression for the gauge field one-forms
\begin{align}
&A_1=\frac{(x_{12}^2dx_1^1-x_{12}^1dx_1^2)+k(x_{12}^2 dx_2^1-x_{12}^1dx_2^2)}{2|\vec{x}_{12}|(|\vec{x}_{12}|+x_{12}^3)}+\frac{x_1^2dx_1^1-x_1^1dx_1^2}{2|\vec{x}_1|(|\vec{x}_1|+x_1^3)}\,,\nonumber\\
&A_2=\frac{k(x_{12}^2 dx_1^1-x_{12}^1dx_1^2)+k^2(x_{12}^2dx_2^1-x_{12}^1dx_2^2)}{2|\vec{x}_{12}|(|\vec{x}_{12}|+x_{12}^3)}+\frac{N_f(x_2^2dx_2^1-x_2^1dx_2^2)}{|\vec{x}_2|(|\vec{x}_2|+x_2^3)}\,,\label{NfsolOmeg}
\end{align}
where we have introduced the shorthand notation
$\vec{x}_{12}=\vec{x}_1+k\vec{x}_2$. This explicitly determines the
metric by inserting (\ref{NfsolOmeg}) into (\ref{metricX8}). However,
due to the complicated form of (\ref{NfsolOmeg}) the complete metric
becomes rather difficult to handle and we therefore will not work with
it directly.

Instead we want to discuss the isometry group of the geometry
(\ref{metricX8}) with solution (\ref{NfsolOmeg}). First of all, there
are two global $U(1)$ symmetries since (\ref{metricX8}) is invariant
under a shift of each of the $\varphi_i$ by a constant. We choose to
parameterize these $U(1)$s in the following manner
\begin{align}
&U(1)_\text{gauge}:\ \left\{\begin{array}{ll}\varphi_1\longmapsto \varphi_1 + \lambda_1 \\ \varphi_2\longmapsto \varphi_2 + \lambda_1\end{array}\right.\,,&&\text{with} &&\lambda_1\in[0,2\pi)\,,\label{U1diag} \\ 
&\nonumber\\
&U(1)_b:\ \left\{\begin{array}{ll}\varphi_1\longmapsto \varphi_1 + \lambda_2 \\ \varphi_2\longmapsto \varphi_2 - \lambda_2\end{array}\right.\, ,&&\text{with} &&\lambda_2\in[0,2\pi)\,,\label{U1glob}
\end{align}
where $\lambda_1$ and $\lambda_2$ are just two constant
parameters.\footnote{Notice that we have called the second $U(1)$
  symmetry $U(1)_b$. This is not by chance, since we will see later on
  that this $U(1)$ maps precisely to the ``baryonic`` $U(1)_b$
  symmetry (\ref{baryonU1}) in the dual gauge theory.} The diagonal
$U(1)_{\text{gauge}}$ can be promoted to a local symmetry provided
that we also transform the gauge potential (\ref{NfsolOmeg}). In fact,
$U(1)_{\text{gauge}}$ is part of a larger $SU(2)_{\text{gauge}}$
``gauge'' symmetry,\footnote{This symmetry is local w.r.t.~the {\em
    internal} coordinates $\vec x_1, \vec x_2$ and therefore global
  w.r.t.~the spacetime coordinates $X_{0,1,2}$.}
 which acts in the usual way on (\ref{NfsolOmeg})
\begin{align}
SU(2)_{\text{gauge}}:\,\varphi_i\longmapsto \varphi_i +\Lambda (\vec{x}_1,\vec{x}_2)\,,&&\text{and} &&A_i\longmapsto A_i-\partial_i \Lambda (\vec{x}_1,\vec{x}_2)\, .\label{SU2gauge}
\end{align}
So far we have only been considering invariances of (\ref{metricX8})
involving $\varphi_i$ and $A_i$, while there is additionally also an
$SO(3)$ symmetry which acts diagonally on $\vec{x}_1$ and $\vec{x}_2$.
As we can see\footnote{For a similar discussion in the context of the
  Taub-Nut space see e.g. \cite{Cotaescu:2003gx}.} from
(\ref{NfsolOmeg}), in order to keep the metric (\ref{metricX8})
invariant, such an $SO(3)$ rotation will only close up to a gauge
transformation of (\ref{NfsolOmeg}). We can, for example, for
$\Omega\in SO(3)$ write a transformation which will leave the metric
invariant in the following manner
\begin{align}
SO(3):\,\left\{\begin{array}{ccl}(\vec{x}_1,\vec{x}_2) & \longmapsto & (\Omega \vec{x}_1,\Omega \vec{x}_2) \\ A_i & \longmapsto & A_i-\partial_i h(\Omega,\vec{x}_1,\vec{x}_2) \\ \varphi_i & \longmapsto & \varphi_i+h(\Omega,\vec{x}_1,\vec{x}_2)\end{array}\right.\,.\label{SO3rotat}
\end{align}
Here, according to \cite{Cotaescu:2003gx}, $h$ is a function of
$\Omega$ and $\vec{x}_{1,2}$, which needs to satisfy
\begin{align}
&h(\Omega_1\Omega_2,\vec{x}_1,\vec{x}_2)=h(\Omega_1,\Omega_2\vec{x}_1,\Omega_2\vec{x}_2)+h(\Omega_2,\vec{x}_1,\vec{x}_2)\,, &&\text{and} &&h(\mathbbm{1},\vec{x}_1,\vec{x}_2)=0\,.
\end{align}
We can therefore summarize that the complete isometry group of the
near-horizon geometry is given by
\begin{align}
SO(3)\times SU(2)_{\text{gauge}}\times U(1)_b\, .\label{isometries}
\end{align}
This fits nicely with our analysis of the symmetries present in the
field theory (see section~\ref{secft}). Indeed, the $SO(3)$ symmetry
takes over the role of the $SU(2)$ R-symmetry group, while we can
identify $SU(2)_{\text{gauge}}$ with the additional global $SU(2)_D$
symmetry present in the dual CFT. As we have already remarked, the
global $U(1)_b$ is identified with $U(1)_b$ of (\ref{baryonU1}). We
should finally also mention that in the limit $N_f=0$ the isometry
group (\ref{isometries}) is in fact enhanced. Most prominently,
$SO(3)$, which in (\ref{SO3rotat}) acts diagonally on
$(\vec{x}_1,\vec{x}_2)$, is enhanced to $SO(3)\times SO(3)$ acting
separately on $\vec{x}_1$ and $\vec{x}_2$. This in turn means that
(\ref{isometries}) becomes isomorphic to $SU(4)\times U(1)_b$, which
ties in nicely with the analysis of the symmetries in the dual
ABJM-model (see \cite{ABJM}). There it was found, that the
three-dimensional Chern-Simons matter theory has an $SU(4)_R$ R-symmetry
and an additional global baryonic $U(1)$.

%%%%%%%%%%%%%%%%%%%%%%%%%%%%%%%%%%%%%%%%%%%%%%%%%%%%%%%%%%%%%%%%%%
%%%%%%%%%%%%%%%%%%%%%%%%%%%%%%%%%%%%%%%%%%%%%%%%%%%%%%%%%%%%%%%%%%
\setcounter{equation}{0}
\section{D6-branes in $AdS_4 \times \cp$} \label{secprobe}

In the previous section we discussed the fully backreacted solution of
the dual gravitational theory. The structure of the corresponding
near-horizon geometry is quite involved, which makes a full
discussion of the supergravity fluctuations of this background technically difficult. A
simpler approach towards a gravity dual of the ABJM theory with
flavour is to treat the fundamental fields in the quenched
approximation. On the gravity side this corresponds to taking the {\em
  probe} limit \cite{KarchKatz}, in which it is assumed that for a
small number of flavours the backreaction of the 
D6-branes may be ignored.  We will therefore embed probe
D6-branes into the (type-IIA) near-horizon geometry of the ABJM model,
which is $AdS_4 \times \cp$ \cite{ABJM}.  Since flavour branes are
spacetime-filling, the D6-branes extend along all the directions of
the $AdS_4$ space and wrap a Lagrangian-(codimension three)-cycle
inside $\cp$.  For consistency of the probe approximation, we need to
make sure that this cycle is stable and supersymmetric.\footnote{The
  analogue in Klebanov-Witten theory with flavour corresponds to the
  embedding of probe D7-branes in $AdS_5 \times T^{1,1}$ studied in
  \cite{Kuperstein, Ouyang, Arean}.}

The most natural guess for such a Lagrangian subcycle is a real
projective space $\rp\subset\cp$. This is due to the well-known
mathematical fact that a $\rp$ is among the simplest codimension-three
cycles inside $\cp$ \cite{Chiang}. Moreover, it is
also known \cite{Oh} that $\rp$ fulfills certain mathematical
stability-criteria under special Hamiltonian deformations. This
already points towards the fact that $\rp$ is indeed a stable cycle
for the probe D6-brane to wrap.

In the following we will not only show that the configuration of a D6
wrapping a $\rp\subset\cp$ is stable but moreover also supersymmetric.
We will proof stability by showing that this cycle gives rise to a
generalized calibration 3-form. As a by-product of this computation
we will find explicit expressions for the Killing spinors of $\cp$.

%%%%%%%%%%%%%%%%%%%%%%%%%%%%%%%%%%%%%%%%%%%%%%%%%%%%%%%%%%%%%%%%%%%%%%%%%%%

\subsection{Geometry of $\cp$}
\subsubsection{Metric and curvature}
Let us start by reviewing some basic facts about the $AdS_4\times \cp$
supergravity solution. According to \cite{ABJM} the metric, the
dilaton and the 2- and 4-form field-strengths are given by
\begin{align}
&ds^2=\frac{R^3}{k}\left(\frac{1}{4}ds_{Ads_4}^2+ds_{\cp}^2\right),
&&e^{2\phi}=\frac{R^3}{k^3}\,, \nonumber\\
&F^{(2)}_{mn}=k J_{mn}\,,
&& F^{(4)}_{\mu\nu\rho\tau}=\frac{3R^3}{8}\epsilon_{\mu\nu\rho\tau}\,.
\end{align}
Here we have used Greek indices $\mu,\nu=1,\ldots,4$ to denote the
directions of $AdS_4$ and Latin indices $m,n=1,\ldots,6$ for the
$\cp$. Moreover, $ds_{\cp}^2$ is the standard Fubini-Study metric
given by
\begin{align}
ds_{\cp}^2=\frac{d\bar{\zeta}_\alpha d\zeta^\alpha}
{(1+\bar{\zeta}_\gamma\zeta^\gamma)^2}+
\frac{\zeta^\alpha\bar{\zeta}_{\beta} d\bar{\zeta}_{\alpha}d\zeta^\beta}
{(1+\bar{\zeta}_\gamma\zeta^\gamma)^4}\,,\label{FubiniStudyGeneral}
\end{align}
with 
\begin{align}
&\zeta_1=\tan\mu\sin\alpha\sin\frac{\vartheta}{2}e^{\frac{i}{2}
(\psi-\varphi+\chi)}\,,
\nonumber\\
&\zeta_2=\tan\mu\cos\alpha e^{\frac{i}{2}\chi}\,,
\nonumber\\
&\zeta_3=\tan\mu\sin\alpha\cos\frac{\vartheta}{2}e^{\frac{i}{2}
(\psi+\varphi+\chi)}\,,\label{coord3}
\end{align}
and $0\leq \mu, \alpha \leq \pi/2$, $0 \leq \vartheta \leq \pi$, $0
\leq \varphi \leq 2\pi$ \cite{PopeWarner}.  More explicitly, in terms
of the left-invariant $SU(2)$ one-forms
\begin{align}
&\sigma_1=\cos\psi d\vartheta+\sin\vartheta\sin\psi d\varphi\,, &&\sigma_2=\sin\psi d\vartheta-\sin\vartheta\cos\psi d\varphi\,, &&\sigma_3=d\psi+\cos\vartheta d\varphi\,,\nonumber
\end{align}
the metric $ds_{\cp}^2$ can  be written as
\begin{align}
ds_{\cp}^2=d\mu^2+\sin^2\mu&\left[d\alpha^2+\frac{1}{4}\sin^2\alpha(\sigma_1^2+\sigma_2^2+\cos^2\alpha\sigma_3^2)+\frac{1}{4}\cos^2\mu(d\chi+\sin^2\alpha\sigma_3)^2\right].
\end{align}
The K\"ahler form $J_{mn}$ is given by
\begin{align}
J&=\frac{1}{2}dA=\frac{1}{4}d\left[\sin^2\mu(d\chi+\sin^2\alpha\sigma_3)\right]
\nonumber\\
&=\sin2\mu (d\chi+\sin^2\alpha(d\psi+\cos\vartheta d\varphi))\wedge
d\mu-\frac{1}{4}\sin^2\mu\sin^2\alpha\sin\vartheta d\varphi\wedge
d\vartheta\,.
\end{align}
For the choice (\ref{FubiniStudyGeneral}) of the metric, $\cp$ is
Einstein satisfying the relation
\begin{align}
R_{mn}=8 g_{mn}\,.\label{Einsteinmetric}
\end{align}
%%%%%%%%%%%%%%%%%%%%%%%%%%

\subsubsection{Lagrangian submanifold}\label{Sect:LagSub}

Given the (complex) parameterization (\ref{coord3}), a $\rp\subset\cp$
is easily found by making sure that $\zeta_{1,2,3}$ all have the same
(fixed) complex phase $\omega$. As we can see by a quick inspection,
this is achieved by solving
\begin{align}
&\psi-\varphi+\chi=\omega, &&\chi=\omega, &&\psi+\varphi+\chi=\omega\,.
\end{align}
For the simplest choice $\omega=0$ the solution $\psi=\varphi=\chi=0$
yields a manifestly real parameterization of $\rp=S^3/\ZZ_2$ with the
standard metric
\begin{align}
ds_{\rp}^2=d\mu^2+\sin^2\mu d\alpha^2+\frac{1}{4}\sin^2\alpha\sin^2\mu d\vartheta^2\,.\label{metricRP}
\end{align}
The isometry group of $\rp$ is $\ZZ_2 \ltimes (SO(3)
\times SO(3))$, where $\ltimes$ denotes a semi-direct product. This means the isometry group
consists of two $SO(3)$ groups and a discrete symmetry exchanging the
two $SO(3)$ groups, see e.g.~\cite{0308022}. The occurrence of two
$SO(3)$'s reflect the $SU(2)$ R-symmetry and the global $SU(2)_D$
symmetry of the field theory.  In the following we will show that a
D6-brane wrapped around this submanifold is indeed a stable and
supersymmetric configuration.

%%%%%%%%%%%%%%%%%%%%%%%%%%%%%%%%%%%%%%%%%%%%%%%%%%%%%%%%%%%%%%%%%%%%%%%%%%%
\subsection{Killing spinors of $\cp$}\label{Sect:KillSpin}

For the construction of a bispinor 3-form in the next subsection, we
will need the Killing spinors of $\cp$. It is a well-known
fact \cite{Nilsson} that there are six Killing spinors on $\cp$. They
can be found as solutions of the following two equations, which stem
from the supervariations of the fermionic degrees of freedom in
supergravity
\begin{align}
&D_m\epsilon-\frac{1}{32}\left(\Gamma_mQ+16\tilde{\Gamma}_m\Gamma_0\right)\epsilon-\frac{9}{16}\Gamma_m\epsilon=0\,,\label{diffkilling}\\
&\frac{3}{8\sqrt{2}}Q\Gamma_0\epsilon+\frac{3}{4\sqrt{2}}\Gamma_0\epsilon=0\,.
\label{eigenspinorrel}
\end{align}
Here we have introduced the quantities
\begin{align}
&Q=J^{mn}\Gamma_{mn}\Gamma_0\,, 
&&\text{and} &&\tilde{\Gamma}_m={J_m}^n\Gamma_n\,.\label{matQ}
\end{align}
Moreover, $\Gamma_m$ are the six-dimensional Dirac matrices. Relation
(\ref{eigenspinorrel}) is an eigenspinor relation for the operator $Q$,
and it was shown in \cite{Nilsson} that $Q$ has the following
eigenvalues
\begin{align}
\{-2,-2,-2,-2,-2,-2,6,6\}\,.
\end{align}
Since we know that the background $AdS_4\times \cp$ preserves $\N=6$
supersymmetry, we conclude that we need to look for eigenspinors to
the eigenvalue $-2$ because the latter has a 6-fold degeneracy. The
most general spinor compatible with this eigenvalue is given by
\begin{align}
\epsilon=\left(\begin{array}{c}f_1+f_2+f_6 \\ -f_6 \\ -f_3+f_4+f_5 \\ f_5 \\ f_4 \\ f_3 \\ f_2 \\ f_1\end{array}\right),\label{killansatz}
\end{align}
where $f_{i=1,\ldots,6}$ are six arbitrary functions of the
coordinates $(\mu,\alpha,\vartheta,\varphi,\psi,\chi)$. The exact
functional dependence can be fixed by inserting this ansatz into the
remaining killing spinor equation (\ref{diffkilling}). This yields a
system of coupled first order partial differential equations, which
can be solved analytically. Since the computations are rather tedious,
we will not relate all the details here, but content ourselves by
giving the explicit solution in appendix
\ref{Sect:KillingSpinors}.

%%%%%%%%%%%%%%%%%%%%%%%%%%%%%%%%%%%%%%%%%%%%%%%%%%%%%%%%%%%%%%%%%%%%%%%%%%%

%%%%%%%%%%%%%%%%%%%%%%%%%%%%%%%%%%%%%%%%%%%%%%%%%%%%%%%%%%%%%%%%%%%%%%%%%%%
\subsection{Embedding of D6-branes}

The Killing spinors of $\cp$ can now be used to compute the
bispinor 3-form
\begin{align}
\Omega_{mnp}=\bar{\epsilon}\Gamma_{mnp}\epsilon \,.
\end{align}
Since the full $\Omega$ is a rather lengthy expression, we refrain from
writing it down completely, but only display the relevant
component. Following our logic in section \ref{Sect:LagSub} concerning
the Lagrangian submanifold, the latter is parameterized by
$(\mu,\alpha,\vartheta)$, while $(\psi,\varphi,\chi)$ are set to
constant values. Therefore, the relevant component of $\Omega$
is $\Omega_{\mu\alpha\vartheta}$ given by
\begin{align}
\Omega_{\mu\alpha\vartheta}=&\frac{1}{2} e^{-\frac{i}{2}
(2 \varphi +\chi +2 \psi )}\left(e^{i (\varphi +\chi +2\psi )}
\lambda_1^2+e^{i\varphi }\lambda_2^2+e^{i \psi }\left(\lambda_4^2+e^{2
i\varphi }\lambda_6^2+e^{i\chi }\left(e^{2
i\varphi}\lambda_3^2+\lambda_5^2\right)\right)\right)\nonumber\\
&\times\sin\alpha \sin^2\mu \,.
\end{align}
Notice that this is a complex expression, which we can separate into
real and imaginary part
\begin{align}
&\text{Re}(\Omega_{\mu\alpha\vartheta})=\frac{1}{2}\sin\alpha\sin^2\mu
\label{REcalibration}\\
&\hspace{0.3cm}\times \left(\left(\lambda_5^2+\lambda_6^2\right)\cos\left(\phi-\frac{\chi}{2}\right)+\left(\lambda_3^2+\lambda_4^2\right)\cos\left(\phi+\frac{\chi}{2}\right)+\left(\lambda_1^2+\lambda_2^2\right)\cos\left(\frac{\chi}{2}+\psi\right)\right),\nonumber\\
&\text{Im}(\Omega_{\mu\alpha\vartheta})=-\frac{1}{2}\sin\alpha\sin^2\mu
\label{IMcalibration}\\
&\hspace{0.3cm}\times\left((\lambda_5^2-\lambda_6^2)\sin\left(\phi-\frac{\chi}{2}\right)+\left(\lambda_4^2-\lambda_3^2\right)\sin\left(\phi
+\frac{\chi}{2}\right)+\left(\lambda_2^2-\lambda_1^2\right)\sin\left(\frac{\chi}{2}+\psi\right)\right). \nonumber
\end{align}
Following \cite{Cascales2004},\footnote{For generalized calibrations
  in backgrounds containing $AdS_4$ see also \cite{Koerber}.} in order
to be a real calibration form the restriction of $\Omega$ to the
Lagrangian submanifold needs to satisfy
\begin{align}
&\text{Im}(\Omega)_{\big|\psi=\varphi=\chi=0}=0\,, &&\text{and} &&\text{Re}(\Omega)_{\big|\psi=\varphi=\chi=0}\simeq\text{Vol}_{\rp}\,, \label{SLcond}
\end{align}
where $\text{Vol}_{\rp}$ is the volume form of the submanifold.
Inserting (\ref{REcalibration}) and (\ref{IMcalibration}) into
(\ref{SLcond}) we indeed find
\begin{align}
&\text{Im}(\Omega)_{\big|\psi=\varphi=\chi=0}=0\,,\\ 
&\text{Re}(\Omega)_{\big|\psi=\varphi=\chi=0}=
\frac{1}{2}(\lambda_1^2+\lambda_2^2+\lambda_3^2
+\lambda_4^2+\lambda_5^2+\lambda_6^2)\sin\alpha\sin^2\mu\,.
\end{align}
Comparing to (\ref{metricRP}) we see that this is indeed proportional
to the volume form of $\rp$ showing that $\Omega$ is indeed a
calibration form. The chosen embedding of the D6-branes is thus stable
and supersymmetric. Naively, we expect that the D6-brane embedding
breaks half of the supersymmetries of the $AdS_4 \times \cp$
background, which ties in with the $\N=3$ supersymmetry of the field
theory. Clearly, to show this precisely would require a careful
analysis of the \mbox{$\kappa$-symmetry}.

%%%%%%%%%%%%%%%%%%%%%%%%%%%%%%%%%%%%%%%%%%%%%%%%%%%%%%%%%%%%%%%%%%%%%%%%%%

\section{Conclusions and open questions}
\setcounter{equation}{0}

In this note we discussed an $\N=3$ version of the ABJM model with
$2N_f$ fields in the fundamental representation of the $U(N)_k \times
U(N)_{-k}$ gauge group.  The theory has a dual description in terms of
$N$ M2-branes at a hypertoric geometry ${\cal M}_8$, given by the
metric (\ref{metricX8}) with (\ref{U}) and (\ref{NfsolOmeg}). We
argued that the isometry of ${\cal M}_8$ is $SU(2) \times SU(2) \times
U(1)$, which matches to the $SU(2)$ R-symmetry of $\N=3$ supersymmetry,
a global $SU(2)_D$ symmetry and a $U(1)_b$ ``baryonic'' symmetry in
the dual field theory. Of course, a complete construction of the
near-horizon geometry would require to also determine the harmonic
function $H$ of the geometry by solving the Laplace equation
(\ref{Laplace}), possibly along the lines of \cite{Hashimoto:2008iv}.

Another approach outlined in this paper is to consider the ``flavour''
D6-branes in the probe approximation, which corresponds to quenched
flavours in the field theory. This requires the embedding of the
D6-branes into the $AdS_4 \times \cp$ near-horizon geometry of the
ABJM model. We showed that $\rp$ is a special Lagrangian three-cycle
in $\cp$ and that D6-branes wrapping $AdS_4 \times \rp$ are stable and
supersymmetric. Since the isometries of $\rp$ match again the global
symmetries of the field theory, we expect that fluctuations of the
D6-branes are dual to bound state operators of (massless)
fundamentals.  It would be interesting to verify this by an explicit
calculation.  This could possibly be done (numerically) using the
Dirac-Born-Infeld action evaluated on the world-volume of the
D6-branes. We leave this for future work.

%%%%%%%%%%%%%%%%%%%%%%%%%%%%%%%%%%%%%%%%%%%%%%%%%%%%%%%%%%%%%%%%%%%%%%%%%%
\section*{Acknowledgements}

We would like to thank Martin Ammon, Johanna Erdmenger, Matthias
Gaberdiel and \mbox{Alfonso} V.~Ramallo for useful discussions and invaluable
comments on a preliminary version of this paper. Moreover, we are
grateful to Stanislav Kuperstein for helpful correspondence. This
research has been supported by the Swiss National Science Foundation.

\vspace{1em}
%%%%%%%%%%%%%%%%%%%%%%%%%%%%%%%%%%%%%%%%%%%%%%%%%%%%%%%%%%%%%%%%%%%%%%%%%%%
\setcounter{equation}{0}
\appendix
\section{Symmetries of the action}\label{Sect:InvCompAct}
In this section we will explore the symmetry content of the action
(\ref{ActPart1})--(\ref{ActPart3}) for the particular values
$\alpha_1=-\alpha_2=1$. In the way (\ref{ActPart1})--(\ref{ActPart3})
is written only $\N=2$ supersymmetry is manifest. In order to exhibit
a possible enhancement of the latter we need to get rid of the
auxiliary fields as they are intimately tied to the $\N=2$ superspace
formulation. We therefore need to work out the action in component
formulation. To this end, we recall the Grassmann expansion of all superfields involved. The chiral superfields are of the form
\begin{align}
&A_i=\supai{i}+\sqrt{2}\theta\psi^{(A)}_{i}+\theta^2 F^{(A)}_{i}\,, &&B_i=\supbi{i}+\sqrt{2}\theta\psi^{(B)}_{i}+\theta^2 F^{(B)}_{i}\,,\\
&Q_i^r=\supqi{i}^r+\sqrt{2}\theta\zeta^r_{i}+\theta^2 G^r_i\,, &&\tilde{Q}^r_i=\suptqi{i}^r+\sqrt{2}\theta\tilde{\zeta}^r_{i}+\theta^2 \tilde{G}^r_i\,, 
\end{align}
while the vector superfields have the expansion
\begin{align}
V_i=2i\theta\bar{\theta}\sigma_i+2\theta\gamma^\mu\bar{\theta}A_{\mu,i}+\sqrt{2}i\theta^2\bar{\theta}\bar{\chi}_i-\sqrt{2}i\bar{\theta}^2\theta\chi_i+\theta^2\bar{\theta}^2D_i  \quad(i=1,2)\,.
\end{align}
Although we have also explicitly written down fermionic components in the Grassmann expansion, we will in the following only focus on the scalar fields, namely $(\supai{i},\supbi{i},\supqi{1,2}^r,\suptqi{1,2}^r)$.\footnote{The calculations for all other components follow in exactly the same manner. However, since the calculation is rather tedious we will not discuss them.} It is then straight-forward to eliminate the auxiliary fields $(F^{(A,B)}_i,G^r_{1,2},\tilde{G}^r_{1,2},\sigma_i,D_i)$ from the action (\ref{ActPart1})--(\ref{ActPart3}), after which the pure scalar part becomes
%{\allowdisplaybreaks
\begin{align}
S=\frac{4\pi^2}{3k^2}\big[&\dq{1}{a}\bdq{1}{a}\dq{1}{b}\bdq{1}{b}\dq{1}{c}\bdq{1}{c}+\dq{2}{a}\bdq{2}{a}\dq{2}{b}\bdq{2}{b}\dq{2}{c}\bdq{2}{c}-4\dq{1}{a}\bdq{1}{b}\dq{1}{c}\bdq{1}{a}\dq{1}{b}\bdq{1}{c}-4\dq{2}{a}\bdq{2}{b}\dq{2}{c}\bdq{2}{a}\dq{2}{b}\bdq{2}{c}+\nonumber\\
%%%%%%%%%%%%%%%%%%%%%%%%
+&\dua{i}{a}\bdua{i}{a}\dua{j}{b}\bdua{j}{b}\dua{k}{c}\bdua{k}{c}+\bdua{i}{a}\dua{i}{a}\bdua{j}{b}\dua{j}{b}\bdua{k}{c}\dua{k}{c}+4\dua{i}{a}\bdua{j}{b}\dua{k}{c}\bdua{i}{a}\dua{j}{b}\bdua{k}{c}-6\dua{i}{a}\bdua{j}{b}\dua{j}{b}\bdua{i}{a}\dua{k}{c}\bdua{k}{c}\nonumber\\
%%%%%%%%%%%%%%%%%%%%5
+&3\dua{i}{a}\bdua{i}{a}\dua{j}{b}\bdua{j}{b}\dq{1}{c}\bdq{1}{c}+3\bdua{i}{a}\dua{i}{a}\bdua{j}{b}\dua{j}{b}\bdq{2}{c}\dq{2}{c}-6\dua{i}{a}\bdua{j}{b}\dua{j}{b}\bdua{i}{a}\dq{1}{c}\bdq{1}{c}-6\bdua{i}{a}\dua{j}{b}\bdua{j}{b}\dua{i}{a}\bdq{2}{c}\dq{2}{c}\nonumber\\
%%%%%%%%%%%%%%%%%%%%%%%%%%%%%%%%%%%%%
+&9\dua{i}{a}\bdua{i}{a}\dq{1}{b}\bdq{1}{b}\dq{1}{c}\bdq{1}{c}+9\bdua{i}{a}\dua{i}{a}\bdq{2}{b}\dq{2}{b}\bdq{2}{c}\dq{2}{c}-6\dua{i}{a}\bdua{i}{a}\dq{1}{b}\bdq{1}{c}\dq{1}{c}\bdq{1}{b}-6\bdua{i}{a}\dua{i}{a}\bdq{2}{b}\dq{2}{c}\bdq{2}{c}\dq{2}{b}\nonumber\\
-&6\dua{i}{a}\bdua{i}{b}\dq{1}{b}\bdq{1}{a}\dq{1}{c}\bdq{1}{c}-6\bdua{i}{a}\dua{i}{b}\bdq{2}{b}\dq{2}{a}\bdq{2}{c}\dq{2}{c}+6\dua{i}{a}\bdua{i}{b}\dq{1}{b}\bdq{1}{c}\dq{1}{c}\bdq{1}{a}+6\bdua{i}{a}\dua{i}{b}\bdq{2}{b}\dq{2}{c}\bdq{2}{c}\dq{2}{a}\nonumber\\
-&6\dua{i}{a}\bdua{i}{b}\dq{1}{c}\bdq{1}{a}\dq{1}{b}\bdq{1}{c}-6\bdua{i}{a}\dua{i}{b}\bdq{2}{c}\dq{2}{a}\bdq{2}{b}\dq{2}{c}-6\dua{i}{a}\bdua{i}{b}\dq{1}{c}\bdq{1}{c}\dq{1}{b}\bdq{1}{a}-6\bdua{i}{b}\dua{i}{a}\bdq{2}{c}\dq{2}{c}\bdq{2}{a}\dq{2}{b}-\nonumber\\
%%%%%%%%%%%%%%%%%%%%%%%%%%%%%%%%%%%%%%%
-&6\bdua{i}{a}\dq{1}{b}\bdq{1}{b}\dua{i}{a}\bdq{2}{c}\dq{2}{c}+12\bdua{i}{a}\dq{1}{b}\bdq{1}{c}\dua{i}{a}\bdq{2}{b}\dq{2}{c}+12\epsilon_{ij}\epsilon^{kl}\dua{i}{c}\bdua{k}{b}\dua{j}{a}\bdua{l}{c}\dq{1}{b}\bdq{1}{a}\nonumber\\
+&12\epsilon^{ij}\epsilon_{kl}\bdua{i}{c}\dua{k}{b}\bdua{j}{a}\dua{l}{c}\bdq{2}{b}\dq{2}{a}\big]\,,\label{FullCovariantAction}
\end{align}%} 
where flavour indices are suppressed. Here we have
arranged all fields in the following doublet form
\begin{align}
&\dua{i}{a}=\left(\begin{array}{c} a_i \\ \bar{b}_i \end{array} \right)\,,&&\text{and} && \dq{1}{a}=\left(\begin{array}{c} q_1 \\ \bar{\tilde{q}}_1 \end{array}\right)\,, &&\text{and} && \dq{2}{a}=\left(\begin{array}{c} \tilde{q}_2 \\ \bar{q}_2 \end{array}\right)\, ,
\end{align}
which exhibits invariance of (\ref{FullCovariantAction}) under two
types of $SU(2)$ symmetries. First of all, we find a $SU(2)_R$
R-symmetry, which acts on the indices $a,b=1,2$. This symmetry acts on
the bifundamental fields $\dua{i}{a}$ as well as on the flavours
$\dq{1,2}{a}$. Apart from this, there is yet another $SU(2)$, which
acts on the index $i,j=1,2$. We call this symmetry $SU(2)_D$ as it can
be understood to be the diagonal of the $SU(2)_A\times SU(2)_B$ global
symmetry, which has already been described in \cite{ABJM}. It is,
however, important to notice that in contrast to the pure ABJM-case
this $SU(2)_D$ commutes with the R-symmetry group $SU(2)_R$ and
therefore does not lead to an additional enhancement of the $SU(2)_R$
R-symmetry group.

As a side remark, we note that the first four lines of
(\ref{FullCovariantAction}) are invariant under a larger $SU(2)_a
\times SU(2)_q$ symmetry rotating separately $a_a^i$ and $q^{1,2}_a$.
However, the last four lines of (\ref{FullCovariantAction}) contain
contractions between the $a_a^i$ and $q^i_a$ fields breaking $SU(2)_a
\times SU(2)_q$ down to $SU(2)_R$.

%%%%%%%%%%%%%%%%%%%%%%%%%%%%%%%%%%%%%%%%%%%%%%%%%%%%%%%%%%%%%%%%%%%%%%%%%%%
%%%%%%%%%%%%%%%%%%%%%%%%%%%%%%%%%%%%%%%%%%%%%%%%%%%%%%%%%%%%%%%%%%%%%%%%%%%
\setcounter{equation}{0}
\section{Killing spinors of $\cp$}\label{Sect:KillingSpinors}
According to the reasoning of section \ref{Sect:KillSpin}, we expect
to find a six-dimensional solution space to the Killing spinor
equation, which is parameterized by the integration constants
$\lambda_1,\ldots, \lambda_6$. With these parameters, we can state the
final result for the general Killing spinor defined in
(\ref{killansatz}): {\allowdisplaybreaks
\begin{align}
f_1&=\frac{1}{2}e^{-\frac{i}{4} (2 \varphi +\chi +2 \psi )}\bigg[e^{\frac{i}{2} (\varphi +\chi +2 \psi )} {\lambda_5} (\cos\alpha+i\cos\mu\sin\alpha)-e^{\frac{i}{2} \varphi} {\lambda_6} \sin\alpha\sin\mu\nonumber\\*
&\hspace{0.5cm}+e^{\frac{i}{2}\psi} \bigg(e^{\frac{i}{2}\chi}\bigg(\cos(\vartheta/2)\left(\left(e^{i\varphi}{\lambda_1}-i{\lambda_3}\cos\alpha\right)\cos\mu+{\lambda_3}\sin\alpha\right)\\*
&\hspace{0.5cm}+\bigg(e^{i \varphi } {\lambda_1} (\sin\alpha-i\cos\alpha\cos\mu)-{\lambda_3}\cos\mu\bigg)\sin(\vartheta/2)\bigg)\nonumber\\*
&\hspace{0.5cm}+\sin\mu\bigg(\left(i{\lambda_2}+e^{i\varphi}{\lambda_4} \cos\alpha\right) \cos (\vartheta/2)+\sin(\vartheta/2)({\lambda_2}\cos\alpha+{\lambda_4}(\sin\varphi-i\cos\varphi ))\bigg)\bigg)\bigg],\nonumber\\
&\nonumber\\
f_2&=\frac{1}{2} e^{-\frac{i}{4} (2 \varphi +\chi +2 \psi )}\bigg[-e^{\frac{i}{2}(\varphi+\chi+2\psi)}{\lambda_5}(\cos\alpha+i\cos\mu\sin\alpha)-e^{\frac{i}{2}\varphi}{\lambda_6}\sin\alpha\sin\mu\nonumber\\*
&\hspace{0.5cm}+e^{\frac{i}{2}\psi}\bigg(e^{\frac{i}{2}\chi}\bigg(\cos(\vartheta/2)\left(\left(e^{i\varphi}{\lambda_1}+i {\lambda_3}\cos\alpha\right)\cos\mu-{\lambda_3}\sin\alpha\right)\nonumber\\*
&\hspace{0.5cm}-\left({\lambda_3}\cos\mu+e^{i\varphi}{\lambda_1}(\sin\alpha-i \cos\alpha\cos\mu)\right)\sin(\vartheta/2)\bigg)\nonumber\\*
&\hspace{0.5cm}+\left(\left(e^{i\varphi}{\lambda_4}\cos\alpha-i{\lambda_2}\right)\cos(\vartheta/2)+\left(ie^{i\varphi}{\lambda_4}+{\lambda_2}\cos\alpha\right)\sin(\vartheta/2)\right)\sin\mu\bigg)\bigg],\\
&\nonumber\\
f_3&=\frac{1}{2} e^{-\frac{i}{2}(\varphi +\psi )}\bigg[e^{\frac{i}{4}(2\varphi-\chi)}{\lambda_6}(\cos\alpha+\cos\mu\sin\alpha)-e^{\frac{i}{4}(2\varphi+\chi+4\psi)}{\lambda_5}\sin\alpha\sin\mu\nonumber\\*
&\hspace{0.5cm}+e^{-\frac{i}{4}(\chi-2\psi)}\bigg(\sin(\vartheta/2)\bigg(\left(-e^{i\varphi}{\lambda_4}-i{\lambda_2} \cos\alpha\right)\cos\mu+{\lambda_2}\sin\alpha\nonumber\\*
&\hspace{0.5cm}+e^{\frac{i}{2}\chi}\left(e^{i\varphi}{\lambda_1}\cos\alpha-i{\lambda_3}\right)\sin\mu\bigg)+\cos(\vartheta/2)\big(e^{i\varphi}{\lambda_4}\sin\alpha\nonumber\\*
&\hspace{0.5cm}+e^{\frac{i}{2}\chi}\left(ie^{i\varphi}{\lambda_1}+{\lambda_3}\cos\alpha\right)\sin\mu+\cos\mu({\lambda_2}-ie^{i\varphi}{\lambda_4}\cos\alpha)\big)\bigg)\bigg],\\
&\nonumber\\
f_4&=\frac{1}{2} e^{-\frac{i}{2}(\varphi+\psi)}\bigg[e^{\frac{i}{4}(2\varphi-\chi)}{\lambda_6}(\cos\alpha+i\cos\mu\sin\alpha)+e^{\frac{i}{4}(2\varphi+\chi+4\psi)}{\lambda_5}\sin\alpha\sin\mu\nonumber\\*
&\hspace{0.5cm}+e^{-\frac{i}{4}(\chi-2\psi)}\bigg(\sin(\vartheta/2)\bigg(\left(e^{i \varphi}{\lambda_4}-i{\lambda_2}\cos\alpha\right)\cos\mu+{\lambda_2}\sin\alpha\nonumber\\*
&\hspace{0.5cm}-e^{\frac{i}{2}\chi}\left(i{\lambda_3}+e^{i\varphi}{\lambda_1}\cos\alpha\right)\sin\mu\bigg)+\cos(\vartheta/2)\bigg(\left(-{\lambda_2}-ie^{i\varphi}{\lambda_4}\cos\alpha\right)\cos\mu\nonumber\\
&\hspace{0.5cm}+e^{i\varphi}{\lambda_4}\sin\alpha-e^{\frac{i}{2}}\sin\mu({\lambda_3}\cos\alpha-ie^{i\varphi}{\lambda_1})\bigg)\bigg)\bigg],\\
&\nonumber\\
f_5&=\frac{1}{2}e^{-\frac{i}{2}(\varphi+\psi)}\bigg[-e^{\frac{i}{4}(2\varphi-\chi)}{\lambda_6}(\cos\alpha-i\cos\mu\sin\alpha)-e^{\frac{i}{4}(2\varphi+\chi+4\psi)}{\lambda_5}\sin\alpha\sin\mu\nonumber\\*
&\hspace{0.5cm}+e^{-\frac{i}{4}(\chi-2\psi)}\bigg(\sin(\vartheta/2)\bigg(\left(-e^{i\varphi}{\lambda_4}-i{\lambda_2}\cos\alpha\right)\cos\mu-{\lambda_2}\sin\alpha\nonumber\\*
&\hspace{0.5cm}+e^{\frac{i}{2}\chi}\left(e^{i\varphi}{\lambda_1}\cos\alpha-i{\lambda_3}\right)\sin\mu\bigg)+\cos(\vartheta/2)\bigg(-e^{i\varphi}{\lambda_4}\sin\alpha\nonumber\\*
&\hspace{0.5cm}+e^{\frac{i}{2}\chi}\left(ie^{i\varphi}{\lambda_1}+{\lambda_3}\cos\alpha\right)\sin\mu+\cos\mu({\lambda_2}-ie^{i\varphi}{\lambda_4}\cos\alpha)\bigg)\bigg)\bigg],\\
&\nonumber\\
f_6&=-\frac{1}{2}e^{-\frac{i}{4}(2\varphi+\chi+2\psi)}\bigg[-e^{\frac{i}{2}(\varphi+\chi+2\psi)}{\lambda_5}(\cos\alpha-i \cos\mu\sin\alpha)-e^{\frac{i}{2}\varphi}{\lambda_6}\sin\alpha\sin\mu\nonumber\\*
&\hspace{0.5cm}+e^{\frac{i}{2}\psi}\bigg(e^{\frac{i}{2}\chi}\big(\cos(\vartheta/2)\left(\left(e^{i\varphi}{\lambda_1}-i{\lambda_3}\cos\alpha\right)\cos\mu-{\lambda_3}\sin\alpha\right)\\*
&\hspace{0.5cm}+\left({\lambda_3}\cos\mu+e^{i\varphi}{\lambda_1}(i\cos\alpha\cos\mu+\sin\alpha)\right)\sin(\vartheta/2)\big)\nonumber\\*
&\hspace{0.5cm}+\sin\mu\left(\left(i{\lambda_2}+e^{i\varphi}{\lambda_4}\cos\alpha\right)\cos(\vartheta/2)+\sin(\vartheta/2)({\lambda_2}\cos\alpha-ie^{i\varphi}{\lambda_4})\right)\bigg)\bigg].\nonumber
\end{align}}

%%%%%%%%%%%%%%%%%%%%%%%%%%%%%%%%%%%%%%%%%%%%%%%%%%%%%%%%%%%%%%%%%%%%%%%
%%%%%%%%%%%%%%%%%%%%%%%%%%%%%%%%%%%%%%%%%%%%%%%%%%%%%%%%%%%%%%%%%%%%%%%
%%%%%%%%%%%%%%%%%%%%%%%%%%%%%%%%%%%%%%%%%%%%%%%%%%%%%%%%%%%%%%%%%%%%%%%

\end{document}